\documentclass[aps,nofootinbib,twocolumn,prd,showpacs,class-pre]{revtex4}
\usepackage{amsmath,amssymb,amsfonts,amsthm}
\usepackage{graphicx}

\usepackage{dsfont}     

\usepackage{subeqnarray}

\newcommand{\be}{\begin{equation}}
\newcommand{\ee}{\end{equation}}
\newcommand{\beq}{\begin{eqnarray}}
\newcommand{\eeq}{\end{eqnarray}}

\newcommand{\bsa}{\begin{subeqnarray}}
\newcommand{\esa}{\end{subeqnarray}}

\newcommand{\I}{{_I}}
\newcommand{\II}{{_{I\!I}}}

\def\boldvec#1{\mbox{\boldmath $#1$\unboldmath}}

\def\lsim{\hbox{ \raise.35ex\rlap{$<$}\lower.6ex\hbox{$\sim$}\ }}
\def\gsim{\hbox{ \raise.35ex\rlap{$>$}\lower.6ex\hbox{$\sim$}\ }}

\newcommand{\bea}{\begin{eqnarray}}
\newcommand{\eea}{\end{eqnarray}}
\newcommand{\beaa}{\begin{eqnarray}}
\newcommand{\eeaa}{\end{eqnarray}}
\newcommand{\ba}{\begin{array}}
\newcommand{\ea}{\end{array}}
\newcommand{\bit}{\begin{itemize}}
\newcommand{\eit}{\end{itemize}}
\newcommand{\ben}{\begin{enumerate}}
\newcommand{\een}{\end{enumerate}}

\def\lab{\label}

\def\lf{\left}

\def\pa{\partial}
\def\ran{\rangle}

\def\ri{\right}

\def\ga{\gamma}
\def\Ga{\Gamma}

\def\ka{\kappa}

\def\om{\omega}
\def\Om{\Omega}
\begin{document}

\noindent KCL-PH-TH/2011-{\bf 19}

\title{Noncommutative spectral geometry, algebra doubling and the
seeds of quantization} \author{Mairi
  Sakellariadou\footnote{mairi.sakellariadou@kcl.ac.uk}}
\affiliation{Department of Physics, King's College, University of
  London, Strand WC2R 2LS, London, U.K.}  \author{Antonio
  Stabile\footnote{anstabile@gmail.com} and Giuseppe
  Vitiello\footnote{vitiello@sa.infn.it}} \affiliation{Facolt\`a di
  Scienze and Gruppo Collegato I.N.F.N., Universit\`a di Salerno,
  I-84100 Salerno, Italy }

\begin{abstract}
A physical interpretation of the two-sheeted space, the most
fundamental ingredient of noncommutative spectral geometry proposed by
Connes as an approach to unification, is presented. It is shown that
the doubling of the algebra is related to dissipation and to the gauge
structure of the theory, the gauge field acting as a reservoir for the
matter field. In a regime of completely deterministic dynamics,
dissipation appears to play a key r\^ole in the quantization of the
theory, according to 't~Hooft's conjecture. It is thus argued that the
noncommutative spectral geometry classical construction carries
implicit in its feature of the doubling of the algebra the seeds of
quantization.
\end{abstract}
\pacs{02.40.Gh, 03.65.Yz, 11.15.-q, 12.10-g}

\maketitle

\section{Introduction and Motivation}
\label{motivation}
Noncommutative Spectral Geometry~\cite{ncg-book1,ncg-book2} (NCSG) is
a rich mathematical theory which combines notions of Noncommutative
Geometry with spectral triples, a mathematical tool conceived by Alain
Connes. Within this context, Connes and collaborators built a model
which offers a purely geometric explanation for the Standard Model
(SM) of electroweak and strong interactions --- the most successful
model of particle physics today at hand --- compatible with
right-handed neutrinos and neutrino masses~\cite{ccm}. This model
succeeds at finding a way to merge the diffeomorphism invariance which
governs General Relativity, with the local gauge invariance which
governs Gauge Theories upon which the SM is based.  The NCSG model has
also been used to derive supersymmetric extensions to the
SM~\cite{Broek:2010jw}.

This unification model lives by construction at high energy scales
(namely at unification scale), thus providing a natural environment to
address unresolved issues of early universe
cosmology~\cite{Nelson:2008uy,Nelson:2009wr,Marcolli:2009in,mmm,Nelson:2010ru,Nelson:2010rt,Sakellariadou:2010nr,Sakellariadou:2011dk}.
Various criticisms have however been raised.  One may for instance
argue that since the model is at present purely classical, strictly
speaking one cannot employ it within the context of the early universe
since then the energy scales were so high that quantum corrections
could no longer be neglected. Or one may oppose that since the action
functional is obtained through a perturbative approach in inverse
powers of the cut-off scale, it ceases to be valid at lower energy
scales relevant for astrophysical studies.  Note that the original
approach may {\sl a priori} also be treated nonperturbatively, however
it is very difficult to compute exactly the spectral action in its
nonperturbative form.  Another criticism that a physicist may have is
that while the model is naturally developed in Euclidean signature,
any physical studies must be performed in Lorentzian signature.

The aim of this paper is twofold: firstly,  to address the
connection between gauge theories and the algebra doubling and offer a
simple physical insight to this rich mathematical theory; secondly, to
reply to some of the above mentioned criticisms.

In what follows, after a short introduction to noncommutative spectral
geometry in Section~II, we show in Section~III how the algebra
doubling, which is an essential mathematical feature of the NCSG
construction, is related to the gauge structure of the theory. We
introduce the notion of dissipation within this context, which in
Section~IV will lead to the quantum aspect of the noncommutative
spectral geometry and the notion of temperature. In our discussion we
will resort to the proposal by
't~Hooft~\cite{'tHooft:1999gk,erice,'tHooft:2006sy}, according which
quantum features and behaviors in a theory would result from a more
fundamental deterministic scenario due to a process of information
loss. In other words, according to 't~Hooft's proposal, quantum
mechanics emerges from an underlying deterministic classical dynamics
acting at an energy scale much higher than the one of our
observations, provided information loss (dissipation) has
occurred. This means that the NCSG ``classical'' construction, holding
at high energy scales, may carry in itself the seeds for quantum
behavior, provided in the same construction there is room for
dissipation.  In the following, we argue that this is indeed the case,
since, as we show, the characterizing feature of the algebra doubling
is related to dissipation, which in turn can be described in terms of
gauge fields.  Thus the two-sheeted space selected in NCSG is related
to gauge theories, as well as to dissipation and to quantization.  We
summarize our physical interpretation of the NCSG purely gravitational
approach to unification in our conclusions.

Before proceeding, it is important to clarify in which sense we
talk of dissipation and of temperature in what follows. This is
necessary because the Standard Model, as is well known, is a zero
temperature Quantum Field Theory (QFT) model describing a closed
(nondissipative) system. It is however equally well known that even in
high energy physics there are many cases in which the zero temperature
approximation is inadequate, as for example in the quark-gluon plasma
physics, unified gauge theory scenarios, astrophysics, cosmology and
in general in all cases one studies critical phenomena in symmetry 
breaking phase transition processes in the early universe (see e.g.,
Ref.~\cite{Kibble}). When we mention temperature we think therefore to
such circumstances where use of thermal field theory is
unavoidable. It is in fact a fortunate case that the doubling of the
algebra in NCSG also implies finite temperature features, as we will
discuss in Sections~IV and VI. We talk of dissipation and open
  systems in the same specific sense one observes that in a system of
  electromagnetically interacting matter field, neither the
  energy-momentum tensor of the matter field, nor that of the gauge
  field, are conserved.  It is, however, $\partial_{\mu} T^{\mu
  \nu}_{\rm matter} = e F^{\mu \nu} j_{\mu} = - \partial_{\mu} T^{\mu
  \nu}_{\rm gauge\ field}$, so that what it is
conserved~\cite{Landau,Leite} is the {\sl total} $T^{\mu \nu}_{\rm
  total} = T^{\mu \nu}_{\rm matter} + T^{\mu \nu}_{\rm gauge
  \ field}$, which is the energy-momentum tensor of the {\it closed}
system \{matter field, electromagnetic field\}: each element of the
couple is {\sl open} (dissipating) on the other one, although the {\sl
  closeness} of the total system is ensured.  It is in this sense that
dissipation enters our discussion of the implications of the algebra
doubling without spoiling the closeness of the SM.


\section{Elements of Noncommutative Spectral Geometry}
\label{NCG}

For the reader's convenience we summarize briefly some of the basic
features and ingredients of NCSG in the present section.

At low energy scales, the laws of physics can be described by the
action functional $S=S_{\rm E-H}+S_{\rm SM}$ which is the sum of the
Einstein-Hilbert action $S_{\rm E-H}$ and the SM action $S_{\rm
  SM}$. While the former depends on the geometry of the underlying
manifold ${\cal M}$, the latter is governed by internal symmetries of
a gauge group. The symmetries of the two parts are also different:
General Relativity is governed by diffeomorphism invariance (outer
automorphisms), while gauge symmetries are based on local gauge
invariance (inner automorphisms).  Near the Planck scale, this sum
fails to capture the correct description of physics, and one may argue
that the distinct feature between the underlying symmetries of the two
parts of $S$ may be at the origin of the unsuccessful search for a
unified theory of all interactions including gravity. The full group
of invariance of the total (including gravity and matter) action
functional $S$ is the semi-direct product ${\cal U}={\cal G}
\rtimes\hbox{\rm Diff}({\cal M})$ of the group ${\cal G}$ of gauge
transformations of the matter sector (the Standard Model) and the
group ${\rm Diff}({\cal M})$ of diffeomorphisms of the manifold ${\cal
  M}$. One thus tries to find a space whose group of diffeomorphisms
is ${\cal U}$; this can be achieved within noncommutative
geometry~\footnote{A main difference between noncommutative spectral
  geometry and other approaches of quantizing gravity is that here one
  is searching for a hidden signature of space-time geometry within
  the functional of gravity coupled to SM at present energy scales,
  instead of postulating the geometry around the Planck scale which
  necessitates an extrapolation by many orders of magnitude.}.

To capture the effect of the SM on the continuous four-dimensional
manifold, Connes considered a model of a two-sheeted space, made from
the product of a four-dimensional smooth compact Riemannian manifold
${\cal M}$ with a fixed spin structure, by a discrete noncommutative
space ${\cal F}$ composed by only two points.  In this approach, the
SM of electroweak and strong interactions is seen as a
phenomenological model, which however specifies the geometry of
spacetime in such a way so that the Maxwell-Dirac action functional
leads to the SM action. Following this proposal, the geometric space
is defined as the tensor product of a continuous geometry ${\cal M}$
for spacetime by an internal geometry ${\cal F}$ for the SM.

An essential step for this proposal is to adapt the notion of metric
for spaces which do not require commutativity of the coordinates. We
will thus sketch the transition from the Riemannian $g_{\mu\nu}$
paradigm, based on the Taylor expansion in local coordinates $x^\mu$
of the square of the line element $ds^2=g_{\mu\nu}dx^\mu dx^\nu$, to
the NCSG paradigm.

The noncommutative nature of the discrete space ${\cal F}$ is given by
the spectral triple $({\cal A, H, D})$, where ${\cal A}$ is an
involution of operators on the finite-dimensional Hilbert space ${\cal
  H}$ of Euclidean fermions, and ${\cal D}$ is a self-adjoint
unbounded operator in ${\cal H}$.  Spectral triples are analogous to
Fourier transform in commutative spaces and are introduced in order to
create a link with experimental data, which are all of a spectral
nature.  It is worth noting that the spectral nature approach is
intrinsic to the noncommutative spectral geometry.

To be more precise, let ${\cal H}=L^2({\cal M},S)$ be the Hilbert space of
  square integrable sections of the spinor bundle, ${\cal
    A}=C^\infty({\cal M})$ be the algebra of smooth functions on
  ${\cal M}$ acting on ${\cal H}$ as simple multiplication operators
\be (f\xi)(x)=f(x)\xi(x)~,~\forall f\in C^\infty({\cal M})
\ \mbox{and}\ \forall \ \xi\in L^2({\cal M},S)~,\nonumber \ee
and $D={\partial\hspace{-5pt}\slash}_{\cal
  M}=\sqrt{-1}\gamma^\mu\nabla_\mu^{\rm s}$ (where $\nabla^{\rm
  s}_\mu$ is the spin connection~\footnote{The spin connection
  $\nabla^{\rm s}_\mu$ is expressed in a vierbein $e$. Let $e^a_\mu$
  be defined as $g_{\mu\nu}=e^a_\mu e^b_\nu \delta_{ab}$, so that
  $\gamma_a=e^\mu_a\gamma_\mu$ satisfy the anticommutation $\{\gamma_a
    ,\gamma_b\}=2\delta_{ab}$. Setting $\gamma_{ab}=(1/4)[\gamma_a
    ,\gamma_b]$, one then gets
\be
\nabla^{\rm s}_\mu=\partial_\mu+{1\over 2}\omega_\mu^{ab}\gamma_{ab}~.\nonumber
\ee
be the Dirac operator on the spin Riemannian manifold ${\cal M}$.}

The algebra ${\cal A}$, related to the gauge group of local gauge
transformations, is the algebra of coordinates; all information about
space is encoded in ${\cal A}$.  In the product noncommutative space
${\cal M}\times{\cal F}$, the algebra ${\cal A}$ is abelian, whereas
the derivative in the discrete direction is a finite difference
quotient. For spaces whose coordinates do not commute, geometry is
defined by specifying the Dirac operator, thus distance is described
through $D$ and not through the metric tensor $g_{\mu\nu}$. The
familiar geodesic formula
\be
d(x,y)={\rm inf}\int_\gamma ds~,
\ee
where the infimum is taken over all possible paths connecting $x$ to
$y$, used to determine the distance $d(x,y)$ between two points $x$
and $y$ within Riemannian geometry, is then replaced by
\be
d(x,y)={\rm sup}\{|f(x)-f(y)|: f\in {\cal A}, ||[D,f]|| \leq 1\}~.
\ee
Within the noncommutative spectral geometry $D$ plays the r\^ole of
the inverse of the line element $ds$.  The operator $D$ corresponds to
the inverse of the Euclidean propagator of fermions, and is given by
the Yukawa coupling matrix which encodes the masses of the elementary
fermions and the Kobayashi--Maskawa mixing parameters.  In conclusion,
NCSG is given by a spectral triple $({\cal A, H}, D)$, in other words
by an involutive algebra ${\cal A}$ represented as operators in
Hilbert space ${\cal H}$ and the line element $ds=1/D$.

The product geometry is specified by the rules:
\beq \lab{1} {\cal A}={\cal A}_1\otimes {\cal A}_2~,~~~~{\cal H}={\cal
  H}_1\otimes {\cal H}_2~,
\eeq
and hence for ${\cal M}\times {\cal F}$ the rules read:
\beq {\cal A}&=&C^\infty({\cal M})\otimes {\cal A}_{\cal
  F}=C^\infty({\cal M},{\cal A}_{\cal F})~,\nonumber\\
{\cal H}&=&L^2({\cal M},S)\otimes {\cal H}_{\cal F}=L^2({\cal
  M},S\otimes{\cal H}_{\cal F})~,\nonumber\\
D&=&{\partial\hspace{-6pt}\slash}_{\cal
  M}\otimes 1+\gamma_5\otimes D_{\cal F}~;
\eeq
$\gamma_5$ is the chirality operator in the four-dimensional case.

Assuming the algebra ${\cal A}$ constructed in the
geometry ${\cal M}\times {\cal F}$ is symplectic-unitary, it must be
of the form~\cite{Chamseddine:2007ia}
\begin{equation}
\mathcal{A}=M_{a}(\mathbb{H})\oplus M_{k}(\mathbb{C})~,
\end{equation}
with $k=2a$ and $\mathbb{H}$ being the algebra of quaternions, which
has a basis $\{1,i\sigma^\alpha\}$, where $\sigma^\alpha
\ (\alpha=1,2,3)$ are the Pauli matrices.  The field of quaternions
$\mathbb{H}$ plays an important r\^ole in this construction and its
choice remains to be explained. To obtain the SM one assumes
quaternion linearity.  The first possible value for the even number
$k$ is 2, corresponding to a Hilbert space of four fermions, but this
choice is ruled out from the existence of quarks. The next possible
value is $k=4$ leading to the correct number of $k^2=16$ fermions in
each of the three generations.  Note that if new particles are
discovered at the Large Hadron Collider (LHC), one may be able to
accommodate them by considering a higher value for the even number
$k$.

Another basic ingredient of the NCSG approach is to consider the
Dixmier trace --- a noncommutative analogue of integration on a
compact $n$-dimensional Riemannian spin manifold --- as the
fundamental functional to define the action of the theory. The Dixmier
trace is then connected with residues of zeta functions.

The noncommutative spectral geometry model is based upon the spectral
action principle stating that, within the context of a product
noncommutative geometry, the bare bosonic Euclidean action is given by
the trace of the heat kernel associated with the square of the
noncommutative Dirac operator and is of the form
\be {\rm Tr}(f(D/\Lambda))~, \ee
where $f$ is a cut-off function and $\Lambda$ fixes the energy scale;
$D$ and $\Lambda$ have physical dimensions of a mass and there is no
absolute scale on which they can be measured.  This action can be seen
{\sl \`a la} Wilson as the bare action at the mass scale $\Lambda$.
The fermionic term can be included in the action functional by adding
$(1/2)\langle J\psi,D\psi\rangle$, where $J$ is the real structure on
the spectral triple and $\psi$ is a spinor in the Hilbert space ${\cal
  H}$ of the quarks and leptons.

Dealing within noncommutative spaces, one obtains the group ${\cal U}$
of symmetries of gravity and matter as the group of automorphisms, and
the $S=S_{\rm E-H} + S_{\rm SM}$ as the spectral action.  Moreover,
the fermions of the Standard Model provide the Hilbert space of a
spectral triple for the algebra, while the bosons (including the Higgs
boson) are obtained through inner fluctuations of the Dirac operator
of the product geometry.

For the four-dimensional Riemannian geometry, the trace ${\rm
  Tr}(f(D/\Lambda))$ is expressed perturbatively in terms of the
geometrical Seeley-deWitt coefficients $a_n$, which are known for any
second order elliptic differential operator,
as~\cite{sdw-coeff,ac1996,ac1997,nonpert}
\beq\label{asymp-exp} {\rm Tr}(f(D/\Lambda))&\sim&
2\Lambda^4f_4a_0+2\Lambda^2f_2a_2+f_0a_4+\cdots\nonumber
\\&& +\Lambda^{-2k}f_{-2k}a_{4+2k}+\cdots~,
\eeq
where  the
smooth even cut-off function $f$, which decays fast at infinity,
appears through its momenta $f_k$ given by:
\beq \nonumber 
f_0 &\equiv& f(0)\\  \nonumber
f_k &\equiv&\int_0^\infty f(u) u^{k-1}{\rm
  d}u\ \ ,\ \ \mbox{for}\ \ k>0 ~,\nonumber\\ \mbox
    f_{-2k}&=&(-1)^k\frac{k!}{(2k)!} f^{(2k)}(0)~.  \nonumber
\eeq
Moreover, since its Taylor expansion at zero vanishes, the
asymptotic expansion Eq.~(\ref{asymp-exp}) reduces to
\be {\rm Tr}(f(D/\Lambda))\sim
2\Lambda^4f_4a_0+2\Lambda^2f_2a_2+f_0a_4~.  \ee
In conclusion, the cut-off function $f$ plays a r\^ole only through
its three momenta $f_0, f_2, f_4$, which are three real parameters in
the model; they are intimately related to the coupling constants at
unification, the gravitational constant, and the cosmological
constant.  In this four-dimensional Riemannian manifold (one  brane
of the two-sheeted space), the term in $\Lambda^4$ gives a
cosmological term, the term in $\Lambda^2$ gives the Einstein-Hilbert
action functional, and the $\Lambda$-independent term yields the
Yang-Mills action for the gauge fields corresponding to the internal
degrees of freedom of the metric.

The computation of the asymptotic expression for the spectral action
functional results to the full Lagrangian for the Standard Model
minimally coupled to gravity, with neutrino mixing and Majorana mass
terms. Thus, this approach leads to a geometric explanation of the SM;
in particular, the vacuum expectation value of the Higgs field is
related to the noncommutative distance between the two sheets.  The
Higgs field is found to be conformally coupled to the Ricci scalar.
The generalized Einstein-Hilbert action contains in addition a
minimally coupled massless scalar field $\sigma$ related to the
distance $d$ between the two sheets by $d\propto e^{-\sigma(y)}$, with
$y$ an element of the product noncommutative space.


\section{Noncommutative spectral geometry, the algebra doubling and the
gauge structure}
\label{classical}

In Ref.~\cite{ncg-book1} Alain Connes considers the work of Heisenberg
establishing, in the early years of Quantum Mechanics (QM), the {\sl
  matrix mechanics} --- where physical quantities are governed by
noncommutative algebra --- and he discusses how close such a discovery
is to experimental reality and how strict is its relation to the
observed discretization of the energy of the atomic levels and of
angular momentum. In this section our aim is twofold: Firstly, we show
that one central ingredient in NCSG, namely the ``doubling'' of the
algebra ${\cal A} \, \rightarrow \,{\cal A}_1\otimes {\cal A}_2$
acting on the ``doubled" space ${\cal H}={\cal H}_1\otimes {\cal H}_2$
(cf., Eq.~(\ref{1})), is also present in the standard QM formalism of
the density matrix and Wigner function. We then show that the doubling
of the algebra is implicit even in the classical theory when
considering the Brownian motion of a particle, which had so an
important r\^ole in the development of the atomistic view of matter,
and it is related to dissipation.  Secondly, in Subsection~\ref{NCGA}
we show that the doubling of the algebra and the dissipation are
related to the gauge structure of the theory.

For the first part of our discussion, we start by considering the
standard expression~\cite{Feynman:1972a} of the Wigner function
\bea \lab{W} W(p,x,t) && \\
 = \frac{1}{2\pi \hbar}&& \int {\psi^* \left(x -
\frac{1}{2}y,t\right)\psi \left(x + \frac{1}{2}y,t\right)
e^{-i\frac{py}{\hbar}}dy} ~. \nonumber \eea
By putting
\be \lab{1a} x_{\pm}=x\pm \frac{1}{2}y ~,\ee
the associated density matrix is
\be\lab{AA8} W(x_{+},x_{-},t) \equiv \langle x_{+}|\rho (t)|x_{-}\rangle =
\psi^* (x_{-},t)\psi (x_{+},t)~, \ee
and the mean value of a quantum operator $A$ is given by
\bea \nonumber \bar{A}(t)&=& \langle \psi (t)|A|\psi (t)\rangle \\ &=&
\int \! \int \psi^* (x_-,t)\, \langle x_-|A|x_+\rangle \,\psi
(x_+,t)dx_+ dx_- \nonumber \\ &=& \int \! \int \langle x_{+}|\rho
(t)|x_{-}\rangle \langle x_-|A|x_+\rangle dx_+ dx_- ~. \lab{ex(2)}
\eea
In the formalism of the density matrix and the Wigner function, the
coordinate $x(t)$ of a quantum particle is thus split into two coordinates
$x_+(t)$ (going forward in time) and $x_-(t)$ (going backward in
time). The forward  and the backward in time evolution of the
density matrix $W(x_{+},x_{-},t)$ is then described by ``two copies"
of the Schr\"odinger equation:
\be i\hbar {\partial \psi (x_{+},t) \over
\partial t}=H_{+}\psi (x_{+},t)~, \lab{(4a)}
\ee
\be -i\hbar {\partial \psi^* (x_-,t) \over \partial t}=H_-\psi^*
(x_-,t)~, \lab{(4b)}
\ee
respectively, i.e.,
\be i\hbar {\partial \langle x_+|\rho (t)|x_-\rangle \over \partial t}=
H \langle x_+|\rho (t)|x_- \rangle~, \lab{(5a)} \ee
where $H$ is given in terms of the two Hamiltonian operators $H_{\pm}$ as
\be  H \,=\, H_+ -H_- ~. \lab{(5b)} \ee
The connection with Alain Connes' discussion of
spectroscopic experiments, noncommutative matrix algebra, energy level
discretization and the algebra doubling is thus evident: the density
matrix and the Wigner function {\it require} the introduction of a
``doubled" set of coordinates $(x_{\pm}, p_{\pm})$ (or $(x,p_{x})$ and
$(y,p_{y})$) and of their respective algebras. Using the two copies of
the Hamiltonian $H_{\pm }$ operating on the outer product of two
Hilbert spaces ${\cal H}_{+} \otimes {\cal H}_{-}$ has been implicitly
required in QM since the very beginning of the theory. Use of
Eqs.~(\ref{(5a)}), (\ref{(5b)}) shows immediately that the eigenvalues
of $H$ are directly the Bohr transition frequencies $h
\nu_{nm}=E_n-E_m$, which was the first  hint towards an explanation
of spectroscopic structure.

We now show that the need to double the degrees of freedom is implicit
even in the classical theory when considering the Brownian motion. We
closely follow Ref.~\cite{Blasone:1998xt} where the results here
summarized are derived.

We recall that in the classical Brownian theory one has the equation
of motion
\be
m\ddot{x}(t)+\ga \dot{x}(t)=f(t)~, \lab{(22)}
\ee
where $f(t)$ is a random (Gaussian distributed) force obeying
\be
<f(t)f(t^\prime )>_{\rm noise}=2\,\ga \,k_BT\; \delta (t-t^\prime)~. \lab{(23)}
\ee
Equation~(\ref{(22)}) can be derived from a Lagrangian in a canonical
procedure by employing a delta functional classical constraint
representation as a functional integral.  By averaging over the
fluctuating force $f$, one indeed obtains~\cite{Blasone:1998xt}
\bea \lab{(25)}
&{}& <\delta[m\ddot{x}+\ga \dot{x}-f]>_{\rm noise}\\ [2mm]
\nonumber &=&
\int {\cal D}y <\exp[{i\over \hbar}
\int dt \;L_f(\dot{x},\dot{y},x,y)]>_{\rm noise}~,
\eea
where
\be L_f(\dot{x},\dot{y},x,y)= m\dot{x}\dot{y}+ {\ga \over
  2}(x\dot{y}-y\dot{x})+fy~. \lab{(26)} \ee
Note that $\hbar$ is introduced solely for dimensional reasons.
We thus see that the constraint condition at the classical level
introduced a new coordinate $y$, and the standard Euler-Lagrange
equations are obtained, namely
\be
{d\over dt}{\partial L_f\over \partial \dot{y}}=
{\partial L_f\over \partial y} ~ ; \ \ \
{d\over dt}{\partial L_f\over \partial \dot{x}}=
{\partial L_f\over \partial x}~,
\ee
i.e.,
\be
m\ddot{x}+\ga \dot{x}=f ~,\ \  \
m\ddot{y}-\ga \dot{y}=0~. \lab{(27)}
\ee
We remark that the Lagrangian system Eqs.~(\ref{(26)})-(\ref{(27)})
were obtained in a completely classical context in the search aimed to
build up a canonical formalism for dissipative
system~\cite{Bateman:1931a,Morse:1953a,Celeghini:1992yv}. The
$x$-system is an {\it open} system. In order to set up the canonical
formalism it is required to {\it close} the system; this is the r\^ole
of the $y$-system, which is the time-reversed copy of the
$x$-system. The $\{x-y\}$ system is thus a closed system.

We also remark that the exact expression for the imaginary part of the
action  reads~\cite{Srivastava:1995yf,DifettiBook}
\be \label{30c} {\rm Im}{\cal S}[x,y]=
\frac{1}{2\hbar}\int_{t_i}^{t_f}\int_{t_i}^{t_f}dt~ds
\,N(t-s)~y(t)~y(s) ~, \ee
where $N(t-s)$ denotes the quantum noise in the fluctuating random
force given by the Nyquist theorem~\cite{Srivastava:1995yf}.

The meaning of Eq.~(\ref{30c}) is that nonzero $y$ yields an
``unlikely process'' in the classical limit ``$\hbar \rightarrow 0$'',
in view of the large imaginary part of the action. At quantum level,
instead, nonzero $y$ may allow quantum noise effects arising from the
imaginary part of the action~\cite{Srivastava:1995yf}. This sheds some
light on the r\^ole played by the doubled degrees of freedom in the
interplay between classical and quantum.  We thus see that the second
sheet cannot be neglected: in the perturbative approach one may drop
higher order terms in the action functional expansion, since they
correspond to unlikely processes at the classical level. However,
these terms may be responsible for quantum noise corrections
and therefore, in order to not preclude the quantization effects, one
should keep them.

\subsection{The gauge structure}\label{NCGA}

Let us now show how the doubling of the degrees of freedom is
related to the gauge structure of the theory. Our subsequent
discussion will thus unveil the  relation between the
two-sheeted space in the NCSG construction and the gauge structure of
the theory.

We consider the equation of the classical one-dimensional damped
harmonic oscillator
\be
m \ddot x + \gamma \dot x + k x  = 0~, \label{2.1a}
\ee
with time independent $m$, $\gamma$ and $k$, which is a simple
prototype of open systems.

As we have seen, to set up the canonical formalism for open
systems, the doubling of the degrees of freedom is required in such a
way  as to complement the given open system with its
time-reversed image, thus obtaining a globally closed system for which
the Lagrangian formalism is well defined.   The doubling of
the $x$ degree of freedom leads to consider the oscillator in the {\it doubled} $y$ coordinate
\be
m \ddot y - \gamma \dot y + k y  = 0 ~.\label{2.1b}
\ee
The system of the oscillator Eq.~(\ref{2.1a}) and its
time-reversed $(\gamma \rightarrow - \gamma)$ image Eq.~(\ref{2.1b})
is then a closed system described by the Lagrangian density
Eq.~(\ref{(26)}) where we put $f = k x$.  The canonically conjugate
momenta $p_{x}$ and $p_{y}$ can now be introduced as customary in the
Lagrangian formalism: \be p_{x} \equiv {{\partial L}\over{\partial
    \dot x}} = m \dot y - \frac{\gamma}{2} y~~~~,~~ \qquad p_{y}
\equiv {{\partial L}\over{\partial \dot y}} = m \dot x +
\frac{\gamma}{2} x~, \label{(2.3)} \ee and the dynamical variables $\{
x , p_{x} ; y , p_{y} \}$ span the new phase-space.

It is convenient to use the coordinates ${{x_1}(t)}$ and ${{x_2}(t)}$
obtained through the (canonical) transformation
\be x_{1}(t) = \frac{x(t) + y(t)}{\sqrt{2}}~, \qquad x_{2}(t) =
\frac{x(t) - y(t)}{\sqrt{2}}~, \ee
in terms of which the motion equations can be rewritten as
\bsa
m \ddot x_1 + \gamma \dot x_2 + k x_1  &=& 0~, \label{2.16}
\\
[1mm]
m \ddot x_2 + \gamma \dot x_1 + k x_2   &=& 0~,
\esa
and $\, p_{1} = m {\dot x}_{1} + (1/2) \ga {x_2}$ ; $p_{2} = - m
{\dot x}_{2} - (1/2) \ga {x_1} \,$.  The Hamiltonian is then
found to be
\bea H &=& H_1 - H_2\nonumber\\
& =& {1 \over 2m} (p_1 - {\gamma\over
  2}x_2)^2 + {k\over 2} x_1^2 \nonumber \\&& -{1
  \over 2m} (p_2 + {\gamma\over 2}x_1)^2 - {k\over 2}
x_2^2~. \label{2.17} \eea
Following
Refs.~\cite{Tsue:1993nz,Blasone:1996yh,Celeghini:1992a,Celeghini:1993a}
we can now introduce the vector potential as
\be A_i = {B\over 2} \epsilon_{ij} x_j ~~~~(i,j = 1,2)~,\label{2.21}
\ee
with
\be B \equiv {c\over{e}}\gamma ~~,~~{\epsilon}_{ii} = 0~~,~~ {\epsilon}_{12}
  = - {\epsilon}_{21} = 1~.\ee
We realize that $H_i$ (with $i=1,2$) in Eq.~(\ref{2.17}) describe two
particles with opposite charges $e_1 = - e_2 = e$ in the (oscillator)
potential $\Phi \equiv (k/2/e)({x_1}^2 - {x_2}^2) \equiv {\Phi}_1
- {\Phi}_2$ with $ {\Phi}_i \equiv (k/2/e){x_i}^{2}$ and in the
constant magnetic field $\boldvec{B}$ defined as $\boldvec{B}=
\boldvec{\nabla} \times \boldvec{A} = - B \boldvec{{\hat 3}}$, namely:
\bea H &=& {H_1} - {H_2}\nonumber\\
& =& {1 \over 2m} (p_1 - {e_1 \over{c}}{A_1})^2
+ {e_1}{\Phi}_1 \nonumber \\&& - {1 \over 2m} (p_2 + {e_2
  \over{c}} A_2)^2 + {e_2}{\Phi}_2~.\label{2.22} \eea
Using Eq.~(\ref{2.21}) the Lagrangian of the system can be written
in the familiar form \bea L &=& {1 \over 2m} (m{\dot x_1} + {e_1
  \over{c}} A_1)^2 - {1 \over 2m} (m{\dot x_2} + {e_2 \over{c}} A_2)^2
\nonumber\\ && - {e^2\over 2mc^2}({A_1}^2 + {A_2}^2) -
e\Phi \label{2.24i} \nonumber\\ &=& {m \over 2} ({\dot x_1}^2 -
{\dot x_2}^2) +{e\over{c}}( {\dot x}_1 A_1 + {\dot x}_2 A_2) -
e{\Phi}~. \label{2.24} \eea
Remarkably, we have the Lorentzian-like (pseudoeuclidean) metric in
Eq.~(\ref{2.24}) (cf. also Eqs.~(\ref{(5b)}), (\ref{2.22}) and
(\ref{(7)}) below).  The ``minus" sign, not imposed by hand, but
required by the doubling of the degrees of freedom, is crucial in our
derivation (and in the NCSG construction).

In conclusion, the doubled coordinate, e.g., $x_2$ acts as the gauge
field component $A_1$ to which the $x_1$ coordinate is coupled, and
{\sl vice versa}. The energy dissipated by one of the two systems is
gained by the other one and viceversa, in analogy to what happens in
standard electrodynamics as observed at the end of Section~I.  The
interpretation is recovered of the gauge field as the bath or
reservoir in which the system is
embedded~\cite{Celeghini:1992a,Celeghini:1993a}. The gauge structure
thus appears intrinsic to the doubling procedure.

Let us see
then how such a conclusion can be also reached in the case of a
fermion field.

For brevity we discuss the simple case of the massless
fermion~\footnote{Extension to the massive fermion case, the boson
  case and non-Abelian gauge transformation groups is possible, {\sl
    see} Refs.~\cite{Celeghini:1992a,Celeghini:1993a}.} and the U(1)
local gauge transformation group. We will see how in this case the
doubling of the algebra ${\cal A} \, \rightarrow \,{\cal A}_1\otimes
{\cal A}_2$ acting on the outer product space ${\cal H}={\cal
  H}_1\otimes {\cal H}_2$ is related with the gauge structure of the
theory.

We consider the classical (pre-quantum) theory. The Lagrangian
of the massless free Dirac field is:
\be L=-\overline{\psi}\gamma^{\mu}\partial_{\mu}\psi ~.  \label{(1)} \ee
Under the U(1) local gauge transformation,
\be \psi(x)\to \exp{[ig\alpha(x)]}\psi(x)  \label{(2)}~, \ee
$L$ transforms as
\be L \to L^\prime = L-ig \partial^{\mu}\alpha(x)
\overline{\psi}(x)\gamma_{\mu}\psi(x)~. \label{(3)} \ee
It is  well known that  in order to  make $L$ invariant  under the
local  gauge  transformation  Eq.~(\ref{(2)}),  the  coupling  of  the
current  $j_{\mu}=i\overline{\psi}\gamma_{\mu}  \psi$  with the  gauge
vector field $A_{\mu}$ has to be introduced in $L$ in such a way that,
when      $\psi(x)$     transforms     as      in     Eq.~(\ref{(2)}),
$j^{\mu}(x)A_{\mu}(x)$ transforms as
\be j^{\mu}(x)A_{\mu}(x) \to j^{\mu}(x)A_{\mu}(x) +
j^{\mu}(x)\partial_{\mu} \alpha(x) ~, \label{(4)} \ee
i.e.,
\be A_{\mu}(x) \to A_{\mu}(x) +
\partial_{\mu}\alpha(x)~. \label{(5)} \ee
The Lagrangian  $L$ modified by the coupling $g j^{\mu}A_{\mu}$
leads to the lagrangian $L_{\rm g}$ defined as
\be
L_{\rm g} = - \overline{\psi} \gamma^{\mu}\partial_{\mu}\psi+
i g \overline{\psi}\gamma^{\mu}\psi A_{\mu}~,
\label{(6)} \ee
 which is by construction invariant under the U(1) local gauge
transformations Eqs.~(\ref{(2)}), (\ref{(5)}),  namely $L_{\rm g}
\to L_{\rm g}^\prime = L_{\rm g}$. As usual, in order for $A_{\mu}$ to
be a dynamical field, the term $-(1/4) F^{\mu \nu}F_{\mu \nu}$
has to be added to  the modified Lagrangian $L_{\rm g}$. Moreover,
the Lorentz gauge condition
\be \partial^{\mu} A_{\mu}(x) = 0~, \label{(5x)}
\ee
has to be adopted in order to ensure that only transverse modes of the
$A_{\mu}$ field enter physical states. As said in the Introduction,
these are represented by square integrable (spinor) functions in the
Hilbert space ${\cal H} = L^2({\cal M},S)$ where the algebra acts by
multiplication operators. Equation~(\ref{(5x)}) expresses the
restriction to the physical states in ${\cal H}$ where the gauge
constraint is satisfied, which we will denote  by $\langle
\partial^{\mu} A_{\mu}(x) \rangle = 0$, where $\langle ... \rangle$
stands for expectation values in the physical states $\langle {\rm
  phys}| ...|{\rm phys} \rangle$.

Now, let us go back to the Lagrangian Eq.~(\ref{(1)}) for a
classical fermion field and show how the doubling of the fermion
degrees of freedom is related, under convenient constraints, to
the local gauge invariance.

The  field algebra is doubled by introducing
the fermion tilde-field $\tilde{\psi}(x)$.
The tilde-system is a ``copy"
(with the same spectrum and couplings) of the $\psi$-system.
The Lagrangian  is written now as
\be \hat L = L - {\tilde L} = - \overline{\psi}
\gamma^{\mu}\partial_{\mu}\psi + \overline{\tilde {\psi}}
\gamma^{\mu}\partial_{\mu} \tilde{\psi}~. \label{(7)} \ee
We assume, for simplicity, that in $\hat L$ there is no coupling term
of the field ${\psi}(x)$ with the tilde field ${\tilde {\psi}(x)}$.
The Hamiltonian for the system is of the form ${\hat H} = H - {\tilde
  H}$ (to be compared with Eq.~(\ref{(5b)})), which in terms of
creation and annihilation operators of the ${\psi}(x)$ and ${\tilde
  {\psi}(x)}$ fields is given by ${\hat H} = \sum_{\bf k} \hbar\,
\om_{\bf k}(a_{\bf k}^{\dag} a_{\bf k} - \tilde a_{\bf k}^{\dag}
\tilde a_{\bf k})$. Let the zero energy eigenstate of ${\hat H}$ be
denoted by $|0({\theta}) \rangle $~\footnote{In other words,
  $|0(\theta) \rangle $ is the vacuum with respect to the fields
  ${\psi(\theta;x)}$ and $\tilde {\psi}(\theta;x)$ obtained from
  ${\psi(x)}$ and $\tilde {\psi}$(x), respectively, by means of the
  Bogoliubov transformation~\cite{Umezawa:1982nv,DifettiBook}.\\}. The
space of states ${\hat {\cal H}} = {\cal H} \otimes {\tilde {\cal H}}
$ is constructed out of $|0({\theta}) \rangle $ by repeated
applications of creation operators of ${\psi(\theta;x)}$ and ${\tilde
  \psi(\theta;x)}$ and is called the $\theta$-representation $\lbrace
|0({\theta}) \rangle
\rbrace$~\cite{Celeghini:1992a,Celeghini:1993a,DifettiBook}.

In the following we consider the subspace  ${\cal H}_{\theta} \,
{\subset} \, \lbrace |0({\theta}) \rangle \rbrace$ made of all the
states $|a \rangle_{\theta} $, including $|0({\theta}) \rangle $,
such that the {\it $\theta$-state condition}
\be [ a_{\bf k}^{\dag} a_{\bf k} - \tilde a_{\bf k}^{\dag} \tilde
  a_{\bf k} ] |a \rangle_{\theta} = 0 ~, \quad \quad {\rm for ~ any}
\quad {\bf k}, \label{ex(11)}\ee
holds for any $|a \rangle_{\theta} $ in ${\cal H}_{\theta}$. This
condition can be shown to be the realization in ${\cal H}_{\theta}$
of the Lorentz gauge condition
Eq.~(\ref{(5x)})~\footnote{Equation~(\ref{ex(11)}) turns out to be
  equivalent to the Gupta-Bleurer condition in quantum
  electrodynamics~\cite{Celeghini:1992a,Celeghini:1993a,DifettiBook}.}.
We have
\be \langle j_{\mu}(x) \rangle_{\theta} = \langle \tilde j_{\mu}(x)
\rangle_{\theta} , \label{(12)} \ee
where $\langle ... \rangle_{\theta}$ denotes matrix elements in
${\cal H}_{\theta}$. We will denote equalities between matrix
elements in ${\cal H}_{\theta}$, say $\langle A \rangle_{\theta} =
\langle B \rangle_{\theta} $, by $A\cong B $ and call them
$\theta$-{\sl weak equalities}. Since they are
equalities among c-numbers, they are classical equalities.

Now, the key point is that, due to Eq.~(\ref{(12)}), the matrix
elements in ${\cal H}_{\theta}$ of the Lagrangian Eq.~(\ref{(7)}) (as well
as of a more general Lagrangian than the simple one presently
considered) are invariant under the simultaneous local gauge
transformations of $\psi$ and $\tilde \psi$  given by Eq.~(\ref{(2)}) and
\be
\tilde \psi(x) \to  \exp{[ig\alpha(x)]}\,  \tilde
\psi(x) ~, \label{(13)} \ee
respectively, i.e.,
\be \langle \hat L \rangle_{\theta}   \to   \langle  \hat L^\prime
\rangle_{\theta} = \langle \hat L \rangle_{\theta}~,~~{\rm
in}~~{\cal H}_{\theta}, \label{ex(14)} \ee
under the gauge transformations Eqs.~(\ref{(2)}),~(\ref{(13)}).

We thus realize that a crucial r\^ole in the $\theta$-weak gauge
invariance of $\hat L$ under Eqs.~(\ref{(2)}), (\ref{(13)}) is played
by the tilde term $\overline{\tilde {
    \psi}}\gamma^\mu\partial_{\mu}\tilde{\psi}$ since it transforms in
such a way  as to compensate the local gauge transformation of the
$\psi$ kinematic term, i.e.,
\be \overline {\tilde \psi} (x) \gamma^{\mu}\partial_{\mu} \tilde \psi
(x) \to \overline {\tilde \psi} (x) \gamma^{\mu} \partial_{\mu} \tilde
\psi (x) + g \partial^{\mu}\alpha(x) \tilde{j}_{\mu}(x). \label{(15)}
\ee
This suggests  to introduce the vector field $A_{\mu}^\prime$ by
\be gj^{\bar \mu} (x) A_{\bar \mu}^\prime(x)\cong \overline {\tilde \psi}
(x) \gamma^{\bar \mu}
\partial_{\bar \mu} \tilde \psi (x)~,~~~
\bar \mu=0,1,2,3~. \label{(16)} \ee
Here and in the following, the bar over ${\mu}$ means no summation
over repeated indices. Equation~(\ref{(15)}) implies that $A_{\mu}^\prime$
transforms as
\be A_{\mu}^\prime(x) \to A_{\mu}^\prime(x) +
\partial_{\mu}\alpha(x)~,
 \label{ex(17)} \ee
when Eqs.~(\ref{(2)}), (\ref{(13)}) are implemented.  This suggests to
identify, in ${\cal H}_{\theta}$, $A_{\mu}^\prime$ with the
conventional U(1) gauge vector field and to introduce it in the
original Lagrangian through the usual coupling term $ig{\overline
  \psi}\gamma^{\mu}\psi A_{\mu}^\prime$.

We remark that provided we restrict ourselves to the $\theta$-weak
equalities, i.e., to matrix elements in ${\cal H}_{\theta}$, matrix
elements of physical observables, which are solely functions of the
${\psi}(x)$ field, are not changed by Eq.~(\ref{(16)}). Moreover,
observables turn out to be invariant under gauge transformations.
Next, one can show that the conservation laws derivable from $\hat L$,
namely in the simple case of Eq.~(\ref{(7)}) the current conservation
laws:
\be
\partial^{\mu}j_{\mu}(x) = 0~~,~~ \quad \quad
\partial^{\mu}\tilde j_{\mu}(x) = 0~, \label{(18b)} \ee
are also preserved as $\theta$-weak equalities when Eq.~(\ref{(16)}) is
adopted.
One may also show that
\be \label{(29)} \partial^{\nu}F_{\mu\nu}^\prime(x) \cong
-gj_{\mu}(x)~~,~~ \quad \quad \partial^{\nu}F_{\mu\nu}^\prime(x) \cong -g
\tilde j_{\mu}(x)~, \ee
in ${\cal H}_{\theta}$. In the Lorentz gauge, from Eq.~(\ref{(29)})
we also obtain the $\theta$-weak relations
\bea  \label{(30)} \partial^{\mu}A_{\mu}^\prime(x) &\cong& 0~,\nonumber\\
\pa^{2} A_{\mu}^\prime(x) &\cong&
gj_{\mu}(x)~.
\eea
In conclusion, our discussion shows the intrinsic gauge properties of
the ``doubling" procedure: we have obtained that in ${\cal
  H}_{\theta}$ the ``doubled algebra" Lagrangian Eq.~(\ref{(7)}) for
the field $\psi$ and its ``double" $\tilde \psi$ can be substituted by
the Lagrangian:
\be \hat L{\rm _g} \cong - {1\over 4}F^{\prime \mu\nu}
F_{\mu\nu}^\prime - \overline \psi \gamma^{\mu}\partial_{\mu}\psi +
ig{\overline \psi}\gamma^{\mu}\psi A_{\mu}^\prime ~,
\label{(31)} \ee
which is indeed the standard $U(1)$ local gauge invariant Lagrangian
for the fermion field $\psi$.  Remarkably, the tilde-kinematical term
is replaced, in a $\theta$-weak sense, by the gauge field-current
coupling. The second equation in Eq.~(\ref{(29)}), shows that the
variations of the gauge field tensor $F_{\mu \nu}^\prime$ have their
source in the current $\tilde j_{\mu}$, which suggests that the tilde
field plays the r\^ole of a ``reservoir". Such an interpretation in
terms of a reservoir, may thus be extended also to the gauge field
$A_{\mu}^\prime$, which indeed acts in a way to ``compensate" the
changes in the matter field configurations due to the local gauge
freedom.

Finally, in the case an interaction term is present in the Lagrangian
Eq.~(\ref{(7)}), $\hat L_{\rm tot}={\hat L}+{\hat L}_{\rm I},~~~{\hat
  L}_{\rm I}=L_{\rm I}-{\tilde L}_{\rm I}$, the above conclusions still hold
provided ${\cal H}_{\theta}$ is an invariant subspace under the
dynamics described by $\hat L_{\rm tot}$.

We close this Section by remarking that it can be also shown that in
the formalism of the algebra doubling a relevant r\^ole is played by
the noncommutative $q$-deformed Hopf algebra~\cite{Celeghini:1998a},
pointing to a deep physical meaning of the noncommutativity in this
construction. Indeed, the map ${\cal A} \, \rightarrow \,{\cal A}_1
\otimes {\cal A}_2$ in Eq.~(\ref{1}) is just the Hopf coproduct map
${\cal A} \, \rightarrow \, {\cal A} \otimes \mathds{1} + \mathds{1}
\otimes {\cal A} \equiv \, {\cal A}_1 \otimes {\cal A}_2$ which
duplicates the algebra.  The Bogoliubov transformation of ``angle"
$\theta$ relating the fields $\psi (\theta; x)$ and ${\tilde \psi}
(\theta; x)$ to $\psi (x)$ and ${\tilde \psi} (x)$, is known to be
obtained by convenient combinations of the {\it deformed} coproduct
operation of the form $\Delta a^{\dag}_q=a^{\dag}_q\otimes q^{1/2} +
q^{-1/2}\otimes a^{\dag}_q$, where $q \equiv q (\theta)$ is the
deformation parameters and $a^{\dag}_q$ are the creation operators in
the $q$-deformed Hopf algebra~\cite{Celeghini:1998a}. These deformed
coproduct maps are noncommutative and the deformation parameter is
related
to the condensate content of
$|0 (\theta) \rangle$ (constrained by the $\theta$-state condition
Eq.~(\ref{ex(11)})).  In this connection it is interesting to observe
that the $q$-derivative is a finite difference derivative, which has
to be compared with the fact that in the NCSG construction the
derivative in the discrete direction is a finite difference quotient,
as mentioned in Section~\ref{NCG}.

A relevant point is that the deformation parameter {\it labels} the
$\theta$-representations $\{|0 (\theta) \rangle\}$ and, for $\theta
\neq \theta'$, $\{|0 (\theta) \rangle\}$ and $\{|0 (\theta')
\rangle\}$ are unitarily inequivalent representations of the canonical
(anti-)commutation rules. This is a characteristic feature of quantum
field theory~\cite{DifettiBook,Umezawa:1982nv}. Its physical meaning
is that an order parameter exists, which assumes different
$\theta$-dependent values in each of the representations. In other
words, the deformed Hopf algebra structure induces the {\it foliation}
of the whole Hilbert space into physically inequivalent subspaces.


\section{Algera doubling, dissipation and quantization}\label{4}

We have considered till now the doubling of the algebra such as the
one occurring in the NCSG construction and have shown that such a
doubling is related to the gauge structure of the theory. We have done
this by considering essentially classical systems and have mentioned
in several points features of such systems at a quantum level. We have
also stressed that the doubling of the system degrees of freedom, say
$x$, amounts to consider the fact that the system is embedded in some
environment, which is indeed described by the doubled $y$ coordinate.

In a series of papers~\cite{'tHooft:1999gk,erice,'tHooft:2006sy}
't~Hooft has discussed classical, deterministic models and has
conjectured that, provided some specific energy conditions are met and
some constraints are imposed, loss of information might lead to a
quantum evolution. In this section, following
Refs.~\cite{Blasone:2000ew,Blasone:2002hq}, we show that in agreement
with 't~Hooft's conjecture, loss of information (dissipation) in a
regime of completely deterministic dynamics appears to be responsible
of the system's quantum mechanical evolution. Our conjecture is then
that the NCSG classical construction carries {\it implicit} in
its feature of the doubling of the algebra the seeds of quantization.

In order to be specific, we consider the classical damped harmonic
$x$-oscillator described by Eq.~(\ref{2.1a}) and its time--reversed
image, the $y$-oscillator Eq.~(\ref{2.1b}). It is also convenient to
put~\cite{Blasone:1996yh} $ x_1 = r \cosh u $, $ x_2 = r \sinh u$, and
\bea\lab{ca} {\cal C} &=& \frac{1}{4 \Om m}\lf[ \lf(p_1^2  - p_2^2\ri)+
m^2\Om^2 \lf(x_1^2 -  x_2^2\ri)\ri]~, \\
\lab{cc}
J_2&=& \frac{m}{2}\lf[\lf( {\dot x}_1 x_2 - {\dot x}_2
x_1 \ri) - {\Ga} r^2 \ri]
~,
\eea
where ${\cal C}$ is taken to be positive and
$$\Ga = {\ga\over 2 m}~,~ \Om = \sqrt{\frac{1}{m}
(\ka-\frac{\ga^2}{4m})}~,~ \mbox{with}~~ \ka >\frac{\ga^2}{4m}~.$$
Using $z=r^2$ and the canonical transformation:
\begin{eqnarray}
&&q_{1} = \int
\frac{dz \;\; m\Omega}{\sqrt{4 J_{2}^{2} + 4m\Omega {\cal{C}
} z - m^{2}\Omega^{2} z^{2}}}\, ,\nonumber\\
&&q_{2} = 2u + \int \frac{dz}{z} \,
\frac{2J_{2}}{\sqrt{4 J_{2}^{2} + 4m\Omega {\cal{C}}
z - m^{2}\Omega^{2} z^{2}}}\, ,\nonumber\\
&&p_1 = {\cal C}\, , \nonumber\\
&&p_2 = J_2 \, ,
\label{can1}
\end{eqnarray}
the system's Hamiltonian Eq.~(\ref{2.17}) can be rewritten as
\bea \lab{pqham}
H &=& \sum_{i=1}^2p_{i}\, f_{i}(q)\,,
\eea
with $f_1(q)=2\Om$, $f_2(q)=-2\Ga$. Note that $\{q_{i},p_i\} =1$, and
the other Poisson brackets are vanishing.

The Hamiltonian Eq.~(\ref{pqham}) belongs to the class of Hamiltonians
considered by 't~Hooft. There, the $f_{i}(q)$ are nonsingular
functions of the canonical coordinates $q_{i}$ and the equations for
the $q$'s, namely $\dot{q_{i}} = \{q_{i}, H\} = f_{i}(q)$), are
decoupled from the conjugate momenta $p_i$. A complete set of observables, called {\em beables}, then exists, which
Poisson commute at all times.  The meaning of this is that
the system admits a deterministic description even when expressed in
terms of operators acting on some functional space of states
$|\psi\ran$, such as the Hilbert space~\cite{erice}. We stress that
such a description in terms of operators and Hilbert space, does not
imply {\em per se} quantization of the system.  As we will see,
quantization is achieved only as a consequence of dissipation.

Thus we see that $J_2$ and ${\cal C}$ are beables (it can be seen from
the Hamiltonian Eq.~(\ref{pqham}) that $q_1$ and $q_2$ are also beables).
Next we put  $H = H_{\rm \I} - H_{\rm \II}$, with
\bea
&&H_{\rm \I} = \frac{1}{2 \Om {\cal C}} (2 \Om {\cal C}
- \Ga J_2)^2 ~~,~~
H_{\rm \II} = \frac{\Ga^2}{2 \Om {\cal C}} J_2^2\,   \label{split}
\eea
and impose the constraint
\bea
\label{thermalcondition}\quad J_2
|\psi\ran = 0~,
\eea
which defines
physical states and guaranties that $H$ is bounded from below.

Due to the constraint Eq.~(\ref{thermalcondition}) we can then write
\bea \lab{17}
H |\psi\ran= H_{\rm \I} |\psi\ran=  2\Om {\cal C}|\psi\ran
= \left( \frac{1}{2m}p_{r}^{2} + \frac{K}{2}r^{2}\right) |\psi \ran \, ,
\eea
with $K\equiv m \Om^2$. We thus realize that $H_{\rm \I}$ reduces to the
Hamiltonian for the two-dimensional ``isotropic'' (or ``radial'')
harmonic oscillator $\ddot{r} + \Om^2 r =0 $.

The physical states are invariant under time-reversal ($|\psi(t)\ran =
|\psi(-t)\ran$) and periodical with period $\tau = 2\pi/\Omega$.

The generic state $|\psi(t)\ran_{H}$ can be written as
\bea\lab{eqt0} |\psi(t)\ran_{H} = {\hat{T}}\lf[ \exp\lf(
  \frac{i}{\hbar}\int_{t_0}^t 2 \Ga J_2 dt' \ri) \ri]
|\psi(t)\ran_{H_{\rm \I}} ~, \eea
where
${\hat{T}}$ denotes time-ordering and the constant $\hbar$, with
dimension of an action, is needed for dimensional reasons.
The states
$|\psi(t)\ran_{H}$ and $|\psi(t)\ran_{H_{\rm \I}}$ satisfy the equations:
\bea \lab{S1}
i \hbar \frac{d}{dt} |\psi(t)\ran_{H} &=& H \,|\psi(t)\ran_{H}~,
\\ \lab{S2}
i \hbar \frac{d}{dt} |\psi(t)\ran_{H_{\rm \I}} &= &2 \Om
{\cal C} |\psi(t)\ran_{H_{\rm \I}} \, .
\eea
Note that $ H_{\rm \I} = 2 \Om{\cal C} $ has the spectrum ${\cal
  H}^n_{\rm \I}= \hbar \Om n$, $n = 0, \pm 1, \pm 2, ...$; since our
choice has been that ${\cal C}$ is positive, only positive values of
$n$ will be considered.

Let us now exploit the periodicity of the physical states
$|\psi\rangle$.  Following Ref.~\cite{Berry}, one may generally write
\begin{eqnarray}
|\psi(\tau)\ran &= &\exp\left( i\phi
- \frac{i}{\hbar}\int_{0}^{\tau}\langle \psi(t)| H |\psi(t) \rangle dt\right)
|\psi(0)\ran
\nonumber \\
&=&  \exp\left(- i2\pi n\right) | \psi(0)\ran \, ,
\label{eqtH}
\end{eqnarray}
i.e.,
$$\frac{ \langle \psi(\tau)| H |\psi(\tau) \rangle }{\hbar} \tau -
\phi = 2\pi n~~,~~ n = 0, 1, 2, \ldots~.$$  Using $\tau = 2
\pi/\Om$ and $\phi = \alpha \pi$ leads to
\bea\lab{spectrum}
{\cal H}_{\rm \I,e\!f\!f}^n \equiv
\langle \psi_{n}(\tau)| H |\psi_{n}(\tau) \rangle= \hbar
\Om \lf( n + \frac{\alpha}{2} \ri) ~.
\eea
The index $n$ has been introduced to exhibit the $n$ dependence of the
state and the corresponding energy. We see that ${\cal H}_{\rm
  \I,e\!f\!f}^n$ gives the effective $n$th energy level of the
physical system, namely the energy given by ${\cal H}_{\rm \I}^n$
corrected by its interaction with the environment. We conclude that
the dissipation term $J_2$ of the Hamiltonian is responsible for the
zero point ($n = 0$) energy: $E_{0} =(\hbar/2) \Om \alpha$.

We remark that in Quantum Mechanics the zero point energy is formally
due to the nonzero commutator of the canonically conjugate $q$ and $p$
operators: the zero point energy {\it is} the ``signature" of
quantization. Our discussion thus shows that dissipation manifests
itself as ``quantization". In other words, the (zero point) ``quantum
contribution" $E_0$ to the spectrum of physical states signals the
underlying dissipative dynamics.

Let us consider further the dynamical r\^ole of $J_2$.
Using $u(t) = - \Ga t$, Eq.~(\ref{eqt0}) can be rewritten as
\bea\lab{eqt0U} |\psi(t)\ran_{H} =
{\hat T}\lf[ \exp\lf(i \frac{1}{\hbar}
\int_{u(t_0)}^{u(t)} 2 J_2 du'\ri) \ri]
|\psi(t)\ran_{H_\I}, \eea
and we have that
\bea \lab{Su}
-i \hbar \frac{\pa}{\pa u} |\psi(t)\ran_{H} = 2J_{2}
|\psi(t)\ran_{H} \, .
\eea
Thus, $2 J_2$ induces translations in the $u$ variable and in
operatorial notation one can write  $p_{u} =- i \hbar
(\pa/\pa u)$.  Equation~(\ref{thermalcondition}) thus defines families
of physical states, representing stable, periodic trajectories.  Note
that $2 J_{2}$ implements transitions from family to family, according
to Eq.~(\ref{Su}). Equation~(\ref{S1}) can be then rewritten as
\bea \lab{S11}
i \hbar \frac{d}{dt} |\psi(t)\ran_{H} = i \hbar \frac{\pa}{\pa t}
|\psi(t)\ran_{H} + i \hbar \frac{du}{dt}\frac{\pa}{\pa u}
|\psi(t)\ran_{H}\, .
\eea
The contribution to the energy due to dissipation is thus described by
 ``translations" in the $u$ variable.

Consider the defining relation for temperature in thermodynamics  (with $k_{\rm B} = 1$)
\bea\lab{T}
\frac{\pa S}{\pa U} = \frac{1}{T}\,.
\eea
Using $S \equiv (2 J_{2}/\hbar)$ and $U \equiv 2 \Om {\cal C}$,
Eq.~(\ref{pqham}) gives $T = \hbar \Ga$.  Provided  $S$ is
identified with the entropy, $\hbar \Ga$ can be regarded as the
temperature.  Thus,  the ``full Hamiltonian'' Eq.~(\ref{pqham})
plays  the r{\^o}le of the free energy ${\cal F}$, and $2
\Ga J_{2}$ represents  the heat contribution in $H$ (or
$\cal F$).  Note that the statement that $2 J_{2}/\hbar$ behaves as
the entropy is not surprising since it controls the dissipative (thus
irreversible loss of information) part of the dynamics.

It is worth noting that the thermodynamical picture outlined above is
also consistent with the results on the canonical quantization of
open systems in quantum field theory~\cite{Celeghini:1992yv}.

\section{Noncommutative geometry and the dissipative interference phase}
\label{5}

We have seen that doubling of the algebra amounts to consider the
system, its environment and their reciprocal interaction.  The
relation which exists in the NCSG construction between the
doubling of the algebra and the noncommutative geometry, finds a
realization in the relation between dissipation and noncommutative
geometry in the plane of the doubled coordinates $(x_1,x_2)$. The
reason is that dissipation implies the appearance of a ``dissipative
interference phase'', a notion which we will clarify in the present
section.

Although in the following we consider the example of
the damped harmonic oscillator and of its time-reversed image, our
conclusions also apply to more general cases.

Since we will consider paths in the doubled coordinate plane, it is
convenient to work with the $(x_+,x_-)$ coordinates, introduced in
Section \ref{classical} (which slightly differ in their definition
from the $(x_1,x_2)$ coordinates).

We remark that $H$ given by Eq.~(\ref{2.17}) does not change its form
when $x_1, x_2, p_1, p_2$ are replaced by $x_+,x_-, p_+, p_-$,
respectively. The components in the $ (x_+,x_-) $ plane of forward and
backward in time velocity $ v_\pm =\dot{x}_\pm$ are then obtained as
\be v_{\pm }={{\partial  H} \over
\partial p_{\pm }} =\pm \, \frac{1}{m}\lf( p_\pm \mp \frac{\ga}{2}\,
x_\mp \ri) ~, \label{(9)} \ee
and they do not commute
\be [v_+,v_-]=i\hbar \,{\ga\over m^2}~.  \label{(10)} \ee
It is thus impossible to fix these velocities $v_+$ and $v_-$ as being
identical~\cite{Sivasubramanian:2003xy}.
By putting $m v_\pm =\hbar K_\pm$,  Eq.~(\ref{(10)}) gives
\begin{equation}
\left[K_+,K_-\right]=\frac{i \ga}{\hbar } \equiv \frac{i}{L^2}\ ,
\label{DP1}
\end{equation}
and a canonical set of conjugate position coordinates
$ (\xi_+,\xi_-)$ may be defined by $\xi_\pm = \mp L^2K_\mp$ so that
\be
\left[\xi_+,\xi_-  \right] = iL^2. \label{DP2} \ee
The commutation relation Eq.~(\ref{DP2}) 
characterizes the noncommutative geometry in the plane $(x_+,x_-)$.

We now show that an Aharanov--Bohm-type phase interference
can always be associated with the noncommutative $(X,Y)$ plane where
\begin{equation} \label{1noncomm}
[X,Y]=iL^2~;
\end{equation}
$L$ denotes the geometric length scale in the
plane~\cite{Sivasubramanian:2003xy}.

Consider a particle moving in the plane along two paths, $ {\cal P}_1
$ and $ {\cal P}_2 $, starting and finishing at the same point, in a
forward and in a backward direction, respectively. Let $ {\cal A} $
denote the resulting area enclosed by the paths. We will show that the
phase interference $\vartheta$ may be written as
\begin{equation}
\vartheta = \frac{\cal A}{L^2}~. \label{IntPhase}
\end{equation}
A phase space action integral
\begin{equation}
{\cal S}({\cal P})=\int_{\cal P} p_i dq^i~, \label{phase1}
\end{equation}
may be associated with each path $ {\cal P} $ (in phase space) for
motion at fixed energy.  The phase interference $ \vartheta $ between
the two paths $ {\cal P}_1 $ and $ {\cal P}_2 $ is given by the
difference
\begin{equation}
 \vartheta =\frac{1}{\hbar }\int_{{\cal P}_1} p_i dq^i - \frac{1}{\hbar }\int_{{\cal P}_2} p_i
dq^i = \frac{1}{\hbar }\oint_{{\cal P}=\partial \Omega } p_i dq^i ~, \label{phase2}
\end{equation}
with $ {\cal P} $ the closed path going from the initial point to
the final point via path $ {\cal P}_1 $ and returning back to the
initial point via $ {\cal P}_2 $. It constitutes the boundary of a
two-dimensional surface $ \Omega $: $ {\cal P}=\partial \Omega
$. Then, due to Stokes theorem, i.e.,
\begin{equation}
\vartheta = \frac{1}{\hbar }\oint_{{\cal P}=\partial \Omega } p_i
dq^i =\frac{1}{\hbar }\int _\Omega (dp_i \wedge dq^i)~,
\label{phase3}
\end{equation}
the phase interference $ \vartheta $
between two alternative paths turns out to be proportional to the ``area''
${\cal A}$ of the
 surface $ \Omega  $ in phase
space $ (p_1,\ldots ,p_f;q^1,\ldots ,q^f)$.

Equation~(\ref{1noncomm}) in the noncommutative plane can be written as
\begin{equation}
[X,P_X]=i\hbar \ \ \ {\rm where} \ \ \ P_X=\left(\frac{\hbar
Y}{L^2}\right)~, \label{phase4}
\end{equation}
and Eq.~(\ref{phase3}) then reads
\begin{equation}
\vartheta =\frac{1}{\hbar }\int _\Omega (dP_X \wedge dX)
=\frac{1}{L^2}\int _\Omega (dY \wedge dX),
\end{equation}
which proves Eq.~(\ref{IntPhase}), i.e., the quantum phase
interference between two alternative paths in the plane is
determined by the noncommutative length scale
$ L  $ and the enclosed area
$ {\cal A} $.

Notice that the existence of a phase interference  is
connected to the zero point fluctuations in the
coordinates; indeed Eq.~(\ref{1noncomm}) implies a zero point
uncertainty relation
$ (\Delta X) (\Delta Y) \ge L^2/2~. $

For  Eq.~(\ref{DP1}) in the  dissipative case,
i.e.,
\begin{equation}
L^2=\frac{\hbar}{\ga},
\label{DissipationLength}
\end{equation}
we then conclude that, provided $x_+ \neq x_-$, the quantum
dissipative phase interference $\vartheta = {\cal A}/L^2 =
{\cal A} \ga/\hbar$ is associated with the two paths ${\cal P}_1$
and ${\cal P}_2$ in the noncommutative plane.

\section{Conclusions}
\label{concl}

We have considered the implications of the central ingredient in the
NCSG, namely the doubling of the algebra ${\cal A}={\cal A}_1\otimes
{\cal A}_2$ acting on the space ${\cal H} = {\cal H}_1\otimes {\cal
  H}_2$.  Firstly, we have shown that the doubling of the algebra is
related to dissipation (in the sense above specified) and the gauge
field structure.  As a result, the two-sheeted geometry must not be
considered as just a simple almost commutative space, which is the
simplest generalization beyond commutative geometries, but instead,
the construction which can lead to gauge fields, required to explain
the Standard Model.  Secondly, by exploiting 't Hooft's conjecture,
according which loss of information within the framework of completely
deterministic dynamics, might lead to a quantum evolution, we have
argued that dissipation, implied by the algebra doubling, may lead to
quantum features. We have thus suggested that the NCSG classical
construction carries {\it implicit} in the doubling of the algebra the
seeds of quantization.

We have shown that in Alain Connes' two-sheeted construction, the
doubled degree of freedom is associated with ``unlikely processes'' in
the classical limit. Thus, in the perturbative approach one may drop
higher order terms in the expansion, since they correspond to unlikely
processes at the classical level.  However, since the higher order
terms in the expansion are the ones responsible for quantum
corrections, the second sheet cannot be neglected at the classical
level, if one does not want to preclude quantization effects.  Put it
differently, the second sheet --- representing gauge fields --- cannot
be neglected once the universe entered the radiation dominated
era. However, at the Grand Unified Theories scale, when inflation took
place, the effect of gauge fields, in other words the discrete space
of two points, is fairly shielded.

At the end of Section~\ref{classical} we have mentioned that the
deformed Hopf algebra plays a relevant r\^ole in the algebra doubling
and it induces the {\it foliation} of the Hilbert space into
physically inequivalent subspaces. These describe different {\it
  phases} of the system, the ground state associated to each of them
being, as known \cite{Umezawa:1982nv,DifettiBook}, a broken symmetry
vacuum characterized by a different ($\theta$-dependent) value of the
order parameter. Variations in the order parameter (derivatives in the
deformation parameter, or, in the language of Section~\ref{4},
translations in the $u$ parameter classifying 't Hooft families of
states) thus describe phase transitions in the system evolution. In
this connection, it is also interesting to observe that the state
$|0({\theta}) \rangle $ can be shown to be a finite temperature state,
which means that the algebra doubling leads to a thermal field
theory~\cite{DifettiBook,Celeghini:1992yv} (also consistently with the
remarks at the end of Section~\ref{4}). We will not comment more on
this point. However, we remark that in the SM, although born and
formulated as a zero temperature QFT model, one cannot avoid to
consider finite temperature effects, as, for example, in the
quark-gluon plasma studies, when studying phase transition processes
in relation to cosmological scenarios in the early universe evolution,
and/or in discussing unification models.  Thermal aspects appear thus
to be a valuable feature of the NCSG construction and of our
discussion in which they are implicit.



\begin{thebibliography}{10}
%

\bibitem{ncg-book1} A.\ Connes, {\sl Noncommutative Geometry}, Academic
  Press, New York (1994).

\bibitem{ncg-book2}  A.\ Connes and M.~Marcolli, {\sl Noncommutative Geometry,
 Quantum Fields and Motives}, Hindustan Book Agency, India (2008).

\bibitem{ccm} A.~H.~Chamseddine, A.~Connes and M.~Marcolli,
  Adv.\ Theor.\ Math.\ Phys.\  {\bf 11}, 991 (2007)
  [arXiv:hep-th/0610241].

\bibitem{Broek:2010jw}
  T.~v.~d.~Broek and W.~D.~van Suijlekom,
  [arXiv:1003.3788 [hep-th]].

\bibitem{Nelson:2008uy}
  W.~Nelson and M.~Sakellariadou,
  Phys.\ Rev.\ D {\bf 81}, 085038 (2010)
  [arXiv:0812.1657 [hep-th]].

\bibitem{Nelson:2009wr}
  W.~Nelson and M.~Sakellariadou,
  Phys.\ Lett.\  B {\bf 680}, 263 (2009)
  [arXiv:0903.1520 [hep-th]].

\bibitem{Marcolli:2009in}
  M.~Marcolli and E.~Pierpaoli,
  arXiv:0908.3683 [hep-th].

\bibitem{mmm}
M.~Buck, M.~Fairbairn and M.~Sakellariadou,
Phys.\ Rev.\ D {\bf 82} 043509 (2010)
[arXiv:1005.1188 [hep-th]].

\bibitem{Nelson:2010ru}
  W.~Nelson, J.~Ochoa and M.~Sakellariadou,
  Phys.\ Rev.\ Lett.\  {\bf 105} (2010) 101602
  [arXiv:1005.4279 [hep-th]].

\bibitem{Nelson:2010rt}
  W.~Nelson, J.~Ochoa and M.~Sakellariadou,
  Phys.\ Rev.\  D {\bf 82} (2010) 085021
  [arXiv:1005.4276 [hep-th]].

\bibitem{Sakellariadou:2010nr}
  M.~Sakellariadou,
Int.\ J.\ Mod.\ Phys.\ D {\bf 20} 785 (2011)
  [arXiv:1008.5348 [hep-th]].

\bibitem{Sakellariadou:2011dk}
  M.~Sakellariadou,
PoS CNCFG {\bf 2010} 028 (2010)
  [arXiv:1101.2174 [hep-th]].

\bibitem{'tHooft:1999gk}
G.~'t Hooft,
Class.\ Quant.\ Grav.\  {\bf 16} 3263 (1999)
[arXiv:gr-qc/9903084].

\bibitem{erice} G.~'t Hooft, in {\em ``Basics and Highlights of
  Fundamental Physics''}, Erice, (1999) [arXiv:hep-th/0003005].

\bibitem{'tHooft:2006sy} G.~'t Hooft,
J.\ Phys.: Conf.  Ser. {\bf 67} 012015 (2007)
[arXiv:quant-ph/0604008].

\bibitem{Kibble} Y.M. Bunkov and H. Godfrin (Eds), {\it Topological defects and the non-equilibrium dynamics of symmetry-breaking phase transitions}, Kluver Academic Publisher, Dordrecht (2000).

\bibitem{Landau} L.D. Landau and E.M. Lifshitz, The Classical Theory of Fields, Butterworth-Heinemann, Oxford, 1980.

\bibitem{Leite} L. Lopez, Gauge field theory: an introduction, Pergamon Press, Oxford 1981.

\bibitem{Chamseddine:2007ia}
  A.~H.~Chamseddine and A.~Connes,
  Phys.\ Rev.\ Lett.\ {\bf 99}, 191601 (2007)
  [arXiv:0706.3690 [hep-th]].

\bibitem{sdw-coeff}
 A.~H.~Chamseddine and A.~Connes, J.~ Math.~Phys. {\bf 47}, 063504 (2006).

\bibitem{ac1996} A.~H.~Chamseddine and A.~Connes,
Phys.\ Rev.\ Lett.\ {\bf 77}, 4868 (1996).

\bibitem{ac1997} A.~H.~Chamseddine and A.~Connes,
  Comm.\ Math.\ Phys.\ {\bf 186}, 731 (1977).

\bibitem{nonpert}
 A.~H.~Chamseddine and A.~Connes, Comm.\ Math.\ Phys.\ {\bf 293}, 867 (2010)
[arXiv:0812.0165 [hep-th]].

\bibitem{Feynman:1972a} R.P.Feynman, {\it Statistical
Mechanics: A set of Lectures},
W.A. Benjamin, Reading, MA (1972).

\bibitem{Blasone:1998xt}
M.~Blasone, Y.~N.~Srivastava, G.~Vitiello and A.~Widom,
Annals Phys.\ {\bf 267}, 61 (1998).

\bibitem{Bateman:1931a} H.~Bateman, Phys. Rev. {\bf 38}, 815 (1931).

\bibitem{Morse:1953a} P.~M.~Morse and H.~Feshbach,  {\sl  Methods of Theoretical Physics},
McGraw-Hill Book Co. (1953).

\bibitem{Celeghini:1992yv}
E.~Celeghini, M.~Rasetti and G.~Vitiello,
Annals Phys.\  {\bf 215} 156 (1992).


\bibitem{DifettiBook} M.~Blasone, P.~Jizba and G.~Vitiello, {\sl
  Quantum Field Theory and its macroscopic manifestations}, Imperial
  College Press, London, (2011)

\bibitem{Srivastava:1995yf}
Y.~N.~Srivastava, G.~Vitiello and A.~Widom,
Annals Phys.\  {\bf 238}, 200 (1995)
[arXiv:hep-th/9502044].

\bibitem{Tsue:1993nz} Y.~Tsue, A.~Kuriyama and M.~Yamamura,
  Prog.\ Theor.\ Phys.  {\bf 91}, 469 (1994).

\bibitem{Blasone:1996yh}
M.~Blasone, E.~Graziano, O.~K.~Pashaev and G.~Vitiello,
Annals Phys.\  {\bf 252} 115 (1996)
[arXiv:hep-th/9603092].

\bibitem{Celeghini:1992a} E.~Celeghini, E.~Graziano, K.~Nakamura and
  G.~Vitiello, Phys.\ Lett.\ {\bf B285}, 98 (1992).

\bibitem{Celeghini:1993a} E.~Celeghini, E.~Graziano, K.~Nakamura and
  G.~Vitiello, Phys.\ Lett. {\bf B304}, 121 (1993).

\bibitem{Umezawa:1982nv} H.~Umezawa, H.~Matsumoto and M.~Tachiki, {\sl
  Thermo Field Dynamics and condensed states}, North-Holland,
  Amsterdam, The Netherlands (1982).

\bibitem{Celeghini:1998a} E. Celeghini, S. De Martino, S. De Siena,
  A. Iorio, M. Rasetti and G. Vitiello, Phys.\ Lett.\ {\bf A244}, 455
  (1998).

\bibitem{Blasone:2000ew}
M.~Blasone, P.~Jizba and G.~Vitiello,
Phys.\ Lett.\ A {\bf 287} 205 (2001)
[arXiv:hep-th/0007138].

\bibitem{Blasone:2002hq}
M.~Blasone, E.~Celeghini, P.~Jizba and G.~Vitiello,
Phys.\ Lett.\ A {\bf 310}, 393 (2003)
[arXiv:quant-ph/0208012]

\bibitem{Berry}
M.~V.~Berry,
Proc.\ Roy.\ Soc.\ Lond.\  {\bf A392}, 45 (1984).

\bibitem{Sivasubramanian:2003xy} S.~Sivasubramanian, Y.~Srivastava,
  G.~Vitiello and A.~ Widom, Phys. Lett. {\bf A311}, 97 (2003)
  [arXiv:quant-ph/0301005].

\end{thebibliography}
\end{document}